\documentclass{elsart}

\usepackage{epsfig}

\begin{document}

\begin{flushright}
{ASITP-99-17}\\
{SNUTP 99-054}
\end{flushright}

\begin{frontmatter}

\title{QCD sum rule analysis of excited $\Lambda_c$ mass parameter
       }

\author[physics]{Jong-Phil Lee\thanksref{jplee}}
\author[itp]{Chun Liu\thanksref{liu}}
\author[physics]{H.S. Song\thanksref{song}}
\thanks[jplee]{\tt jplee@phya.snu.ac.kr}
\thanks[liu]{\tt liuc@itp.ac.cn}
\thanks[song]{\tt hssong@physs.snu.ac.kr}
\address[physics]{Department of Physics and Center for Theoretical Physics,
  Seoul National University, Seoul 151-742, Korea}
\address[itp]{Institute of Theoretical Physics, Chinese Academy of Sciences, 
  P.O. Box 2735, Beijing 100080, China}

\begin{keyword}
excited heavy baryon,
QCD sum rules,\\
heavy baryon mass 

\begin{abstract}

The mass parameter of orbitally excited $\Lambda_c$ baryons is calculated
by using QCD sum rule in the framework of heavy quark effective theory.
Two kinds of interpolating current for the excited heavy baryons are
introduced.  It is obtained that 
$\bar{\Lambda}=1.08^{+0.095}_{-0.104}$ GeV for the non-derivative current 
and $\bar{\Lambda}=1.06^{+0.090}_{-0.107}$ GeV for the current
with derivative.  These results are consistent with experimental data.

\end{abstract}

\end{keyword}
\end{frontmatter}


\newpage

\section{Introduction}

Study of heavy baryons widens the way to test the Standard Model.
The experimental data on the heavy baryons are accumulating and waiting
for reliable theoretical calculations.  The heavy
quark effective theory (HQET) \cite{Isgur} provides a model-independent
method for analyzing hadrons containing one heavy quark.  It simplifies
the analysis greatly because of the heavy quark symmetry (HQS) and allows
us to expand the physical quantity in powers of $1/m_Q$ systematically,
where $m_Q$ is the heavy quark mass.  To obtain detailed predictions,
however, one needs to combine it with some nonperturbative methods.
\par
QCD sum rule is a powerful nonperturbative method based on QCD
\cite{Shifman}.  It takes into account the nontrivial QCD vacuum,
parametrized by various vacuum condensates.
QCD sum rule has been used successfully to analyze the properties of
light mesons \cite{Shifman} and light baryons \cite{Ioffe}.
Heavy mesons \cite{Neubert,Dai0} and the ground state heavy baryons
\cite{Grozin,Dai,Groote} have been analyzed in the framework of HQET
sum rules with $1/m_Q$ corrections and $\alpha_s$ radiative corrections 
included.
\par
With the discovery of the orbitally excited charmed baryons \cite{PDG}, 
it needs to investigate them theoretically \cite{Falk}.  This
investigation will extend our ability in the application of QCD.  It can
also help us foresee other excited heavy baryons undiscovered yet.
Within the framework of large $N_c$ HQET, we have studied the excited
heavy baryon spectrum \cite{jplee}.  In this paper, we use QCD sum rule
to calculate the mass parameter ${\bar\Lambda}$ of the excited heavy
baryons in the leading order $\alpha_s$ expansion.  ${\bar\Lambda}$ is
defined in the heavy baryon mass expansion,
\begin{equation}
M=m_Q+{\bar\Lambda}+O(\frac{1}{m_Q})~.
\end{equation}
It is independent of the heavy quark spin and flavor, and describes
mainly the contribution of the light degrees of freedom in the baryon.
The quantum  numbers that describe the hadrons are angular momentum $J$ 
and isospin $I$.  For the heavy hadrons, because of HQS, the total 
angular momentum of the light degrees of freedom $J^l$ becomes a good 
quantum number.  In this case, the excited hadron spectrum shows the 
degeneracy of pair of states which are related to each other by HQS.  
For the baryons, the light degrees of freedom look like a collection of
$N_c-1$ light quarks.  There are two kinds of excitation 
\cite{jplee,Cho}.  One is the symmetric representation of the $N_c-1$ 
light quarks in the light quark spin-flavor space.  The other is the 
mixed representation of the light quarks.  In the constituent picture, 
the former kind corresponds to that the two light quarks with vanishing 
relative orbital angular momentum move around the heavy quark with 
orbital angular momentum $L=1$.  In the latter case, one light quark is 
$L=1$ excited while the other light quark and the heavy quark are in 
ground state.  Quark models \cite{Copley} indicate that the former lies 
about 150 MeV below the latter.  In each representation, the states are 
not unique.  The lowest lying pair of states in the symmetric 
representation are denoted as $\Lambda_{c1}({\frac{1}{2}}^-)^+$ and 
$\Lambda_{c1}({\frac{3}{2}}^-)^+$ with "$-$" standing for the parity.  
It is these states that we are going to calculate.  
\par
In Sec. 2, the interpolating fields of the excited heavy baryons are 
constructed.  And the two-point correlation function is calculated.  
Sec. 3 presents the numerical results.  A brief summary and discussion 
are made in Sec. 4.

\section{Interpolating fields and the sum rules}

The heavy baryon current is generally expressed as 
\begin{equation}
j(x)=\epsilon_{ijk}[q^{iT}(x)C\Gamma\tau q^j(x)]\Gamma^\prime h^k_v(x)~,
\end{equation}
where $i,~j,~k$ are the color indices, $C$ is charge conjugation, and
$\tau$ is the isospin matrix.  $q(x)$ is a light quark field, while
$h_v(x)$ is the heavy quark field with velocity $v$.  $\Gamma$ and
$\Gamma^\prime$ are some gamma matrices which describe the structure of
the baryons.  We adopt two kinds of simple forms for $\Gamma$ and
$\Gamma^\prime$.  Usually $\Gamma$ and $\Gamma^\prime$ with least number
of derivatives are used in the QCD sum rule method.  The sum rules then
have better convergence in the high energy region and often have better
stability.  For the ground state heavy baryons, the way to write down
the $\Gamma$ and $\Gamma^\prime$ was discussed in Refs. \cite{Grozin}
and \cite{Dai}.  In our case, the angular momentum and parity of the
light degrees of freedom are $1^-$.  They are the same as that of
$\Sigma_Q({\frac{1}{2}}^+)$ and $\Sigma_Q^*({\frac{3}{2}}^+)$ except the
parity.  Therefore the choice of $\Gamma$ and $\Gamma^\prime$ without
derivatives for $\Lambda_{Q1}({\frac{1}{2}}^-)$ and
$\Lambda_{Q1}({\frac{3}{2}}^-)$ is 
\begin{eqnarray}
\Gamma(\frac{1}{2})=(a+bv\hspace{-1.7mm}/)\gamma^\mu\gamma_5~,~~~~~
&&\Gamma^\prime(\frac{1}{2})=\gamma_{\mu t}\gamma_5~,\nonumber\\
\Gamma(\frac{3}{2})=(a+bv\hspace{-1.7mm}/)\gamma_\nu\gamma_5~,~~~~~
&&\Gamma^\prime(\frac{3}{2})=-g^{\mu\nu}_t
+\frac{1}{3}\gamma^\mu_t\gamma^\nu_t\equiv\Lambda^{\mu\nu}~,
\label{noderiv}
\end{eqnarray}
where a transverse vector $A^\mu_t$ is defined to be
$A^\mu_t\equiv A^\mu-v^\mu v\cdot A$, and 
$g^{\mu\nu}_t=g^{\mu\nu}-v^\mu v^\nu$.  
$a$, $b$ are arbitrary numbers between 0 and 1.
\par
On the other hand, the above discussed constituent picture as well as the
experience from the excited heavy mesons \cite{Dai0}, helps us to choose
the following form of $\Gamma$ and $\Gamma'$ with one derivative for
$\Lambda_{Q1}({\frac{1}{2}}^-)$ and $\Lambda_{Q1}({\frac{3}{2}}^-)$,
\begin{eqnarray}
\Gamma(\frac{1}{2})=(a+bv\hspace{-1.7mm}/)\gamma_5~,~~~~~ 
&&\Gamma^\prime(\frac{1}{2})=
\frac{{\overleftarrow D_t}\hspace{-4.3mm}/}{M}\gamma_5~,\nonumber\\
\Gamma(\frac{3}{2})=(a+bv\hspace{-1.7mm}/)\gamma_5~,~~~~~
&&\Gamma^\prime(\frac{3}{2})=\frac{1}{3M}({\overleftarrow D}^\mu_t
+{\overleftarrow D_t}\hspace{-4.3mm}/~\gamma^\nu_t)~,
\label{deriv}
\end{eqnarray}
where $M$ in Eq. (\ref{deriv}) is some hadronic mass scale to make 
$\Gamma^\prime$ dimensionless.
\par
Furthermore it should be noted that the two possible choices for
$\Gamma$ are accounted through introducing parameters $a$ and $b$.
The baryonic decay constants in the HQET are defined as follows,
\begin{eqnarray}
\langle 0|j|\Lambda_{Q1}(\frac{1}{2})\rangle &=&f_{\frac{1}{2}}u~,
\nonumber\\
\langle 0|j_\mu|\Lambda_{Q1}(\frac{3}{2})\rangle &=&
\frac{1}{\sqrt{3}}f_{\frac{3}{2}}u_\mu
\end{eqnarray}
where $u$ and $u_\mu$ are the spinor and Rarita-Schwinger spinor,
respectively.  $f_{\frac{3}{2}}$ is the same as $f_{\frac{1}{2}}$ in the
heavy quark limit.  The decay constants corresponding to the two choices
of the baryon interpolating currents in Eqs. (\ref{noderiv}) and
(\ref{deriv}) are not related to each other, although they are at the
same order.
\par
\begin{figure}
\vskip 2cm
\begin{center}
\epsfig{file=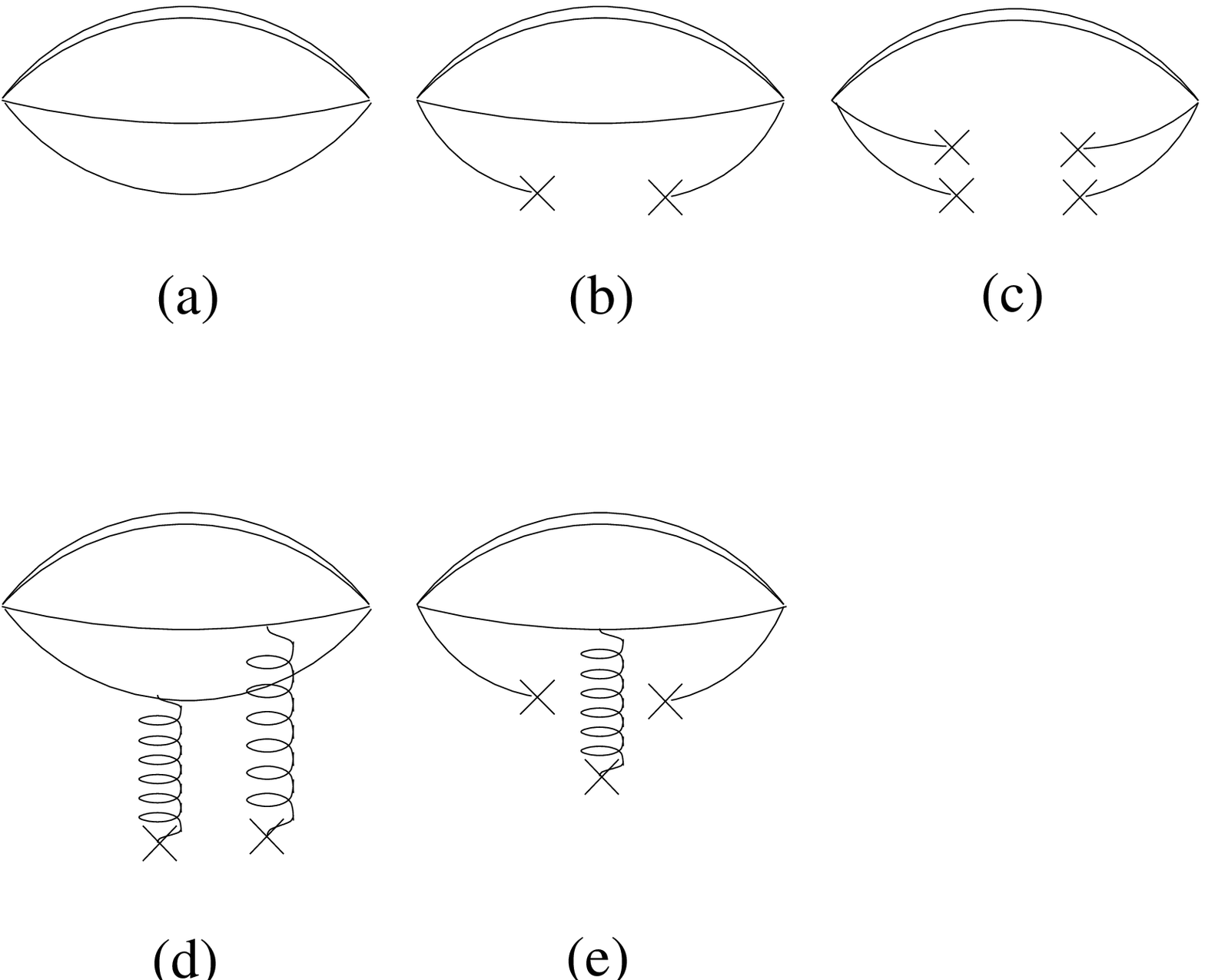, height=10cm}
\end{center}
\caption{}
\label{noderivfig}
\end{figure}
\begin{figure}
\vskip 2cm
\begin{center}
\epsfig{file=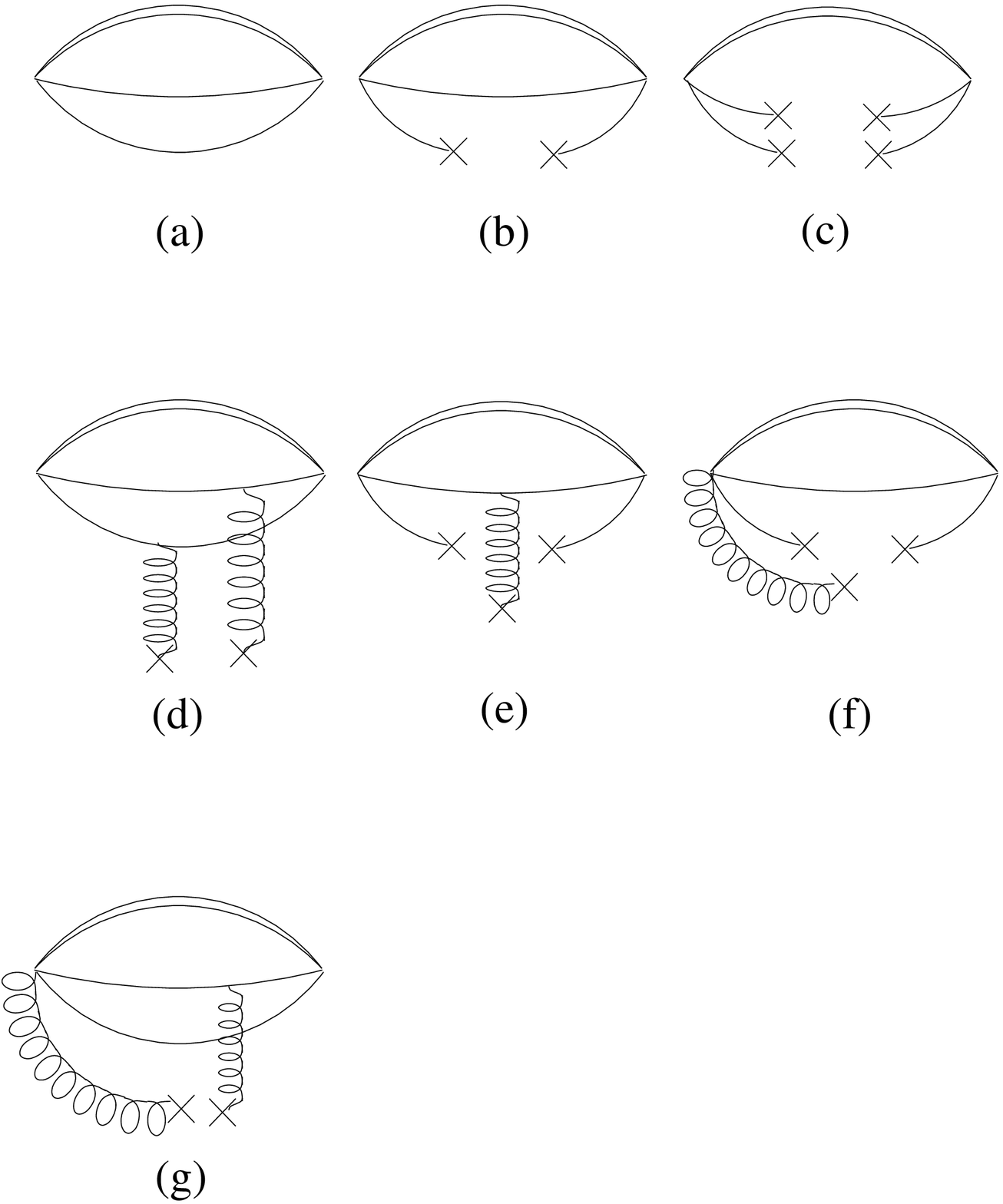, height=15cm}
\end{center}
\caption{}
\label{derivfig}
\end{figure}

For analyzing the masses by QCD sum rule, the following two-point 
correlation function is evaluated,
\begin{equation}
\Pi=i\int d^4 x e^{ikx}\langle0|{\rm T}\{j(x),{\bar j}(0)\}|0\rangle~.
\end{equation}
The hadronic representation of $\Pi$ for $\Lambda_{Q1}({\frac{1}{2}}^-)$
at leading order of $1/m_Q$ is 
\begin{equation}
\Pi=\frac{f^2}{{\bar\Lambda}-\omega}
\frac{1+v\hspace{-1.7mm}/}{2}+{\rm res.}~,
\label{hadronic}
\end{equation}
where $\omega=v\cdot k$.  For $\Lambda_{Q1}({\frac{3}{2}}^-)^+$, $\Pi$ has
the same form except the factor $\frac{1+v\hspace{-1.7mm}/}{2}$ which is
replaced by $\Lambda^{\mu\nu}$.  $\Pi$ can be also calculated in terms of
quark and gluon language with vacuum condensates.  This establishes the
sum rule.  We use the commonly accepted assumption of quark-hadron
duality for the resonance part of Eq. (\ref{hadronic}),
\begin{equation}
{\rm res.}=\frac{1}{\pi}\int_{\omega_c}^{\infty} d\nu
              \frac{{\rm Im}\Pi^{\rm pert}(\nu)}{\nu-\omega}~,
\end{equation}
where $\Pi^{\rm pert}$ is the perturbative contribution to $\Pi$, and
$\omega_c$ is the continuum threshold.  After the Borel transformation
and dropping the common matrix of $\frac{1+v\hspace{-1.7mm}/}{2}$ or
$\Lambda^{\mu\nu}$, we have
\begin{equation}
{\bar\Lambda}=\frac{d}{d(-1/T)}{\rm ln}\Bigg[
  \frac{1}{\pi}\int_0^{\omega_c}{\rm Im}\Pi^{\rm pert}(\nu)e^{-\nu/T}d\nu
  +{\hat B}_T^\omega\Pi^{\rm cond}(\omega)\Bigg]~,
\label{mass}
\end{equation}
where ${\hat B}_T^\omega$ means the Borel transformation, $T$ is the
Borel parameter, and $\Pi^{\rm cond}$ is the condensate contributions to
$\Pi$.  Here the right-handed side must be understood as such that 
$\frac{1+v\hspace{-1.7mm}/}{2}$ or $\Lambda^{\mu\nu}$ is dropped.
Relevant Feynman diagrams contributing to Eq. (\ref{mass}) are listed in
Figs. \ref{noderivfig} and \ref{derivfig} corresponding to the choice of 
Eqs. (\ref{noderiv}) and (\ref{deriv}), respectively.  The fixed point
gauge is used as in \cite{Novikov}.  
For Fig. \ref{noderivfig}, the dependence of ${\bar\Lambda}$ on $a$ and
$b$ is canceled in Eq. (\ref{mass}).  We obtain 
\begin{eqnarray}
\frac{1}{\pi}{\rm Im}\Pi_{1a}(\nu)&=&
 \frac{N_c!}{480\pi^4}\nu^5{\rm Tr}(\tau^\dagger\tau)
 {\rm Tr}(v\hspace{-1.7mm}/\Gamma v\hspace{-1.7mm}/{\tilde\Gamma})
 \Bigg(\Gamma^\prime\frac{1+v\hspace{-1.7mm}/}{2}
 {\tilde\Gamma}^\prime\Bigg)~,\nonumber\\
{\hat B_T^\omega}\Pi_{1b}(\omega)&=&
 \frac{N_c!}{24\pi^2}{\rm Tr}(\tau^\dagger\tau)
 \Big(\langle{\bar q}q\rangle T^3
    -\frac{1}{16}\langle \bar{q}g\sigma\cdot G q\rangle T\Big)
 {\rm Tr}(\Gamma v\hspace{-1.7mm}/{\tilde\Gamma}
     -{\tilde\Gamma}v\hspace{-1.7mm}/\Gamma)\nonumber\\
&&\times
 \Bigg(\Gamma^\prime\frac{1+v\hspace{-1.7mm}/}{2}
 {\tilde\Gamma}^\prime\Bigg)~,\nonumber\\
{\hat B_T^\omega}\Pi_{1c}(\omega)&=&
 -\frac{N_c!}{144}\langle{\bar q}q\rangle ^2 {\rm Tr}(\tau^\dagger\tau)
 {\rm Tr}(\Gamma{\tilde\Gamma})
 \Bigg(\Gamma^\prime\frac{1+v\hspace{-1.7mm}/}{2}
 {\tilde\Gamma}^\prime\Bigg)~,\nonumber\\
{\hat B_T^\omega}\Pi_{1d}(\omega)&=&
 \frac{\langle\alpha GG\rangle}{384\pi^2}T^2{\rm Tr}(\tau^\dagger\tau)
 {\rm Tr}(v\hspace{-1.7mm}/\gamma_5\Gamma v\hspace{-1.7mm}/\gamma_5
 {\tilde\Gamma}-\gamma^\alpha\gamma_5\Gamma\gamma_\alpha\gamma_5
 {\tilde\Gamma})\Bigg(\Gamma^\prime\frac{1+v\hspace{-1.7mm}/}{2}
 {\tilde\Gamma}^\prime\Bigg)~,\nonumber\\
{\hat B_T^\omega}\Pi_{1e}(\omega)&=&
 -\frac{\langle \bar{q}g\sigma\cdot G q\rangle}{1536\pi^2}T{\rm Tr}
 (\tau^\dagger\tau){\rm Tr}(\sigma_{\mu\nu}\Gamma\{\sigma^{\mu\nu},
 v\hspace{-1.7mm}/\}{\tilde\Gamma}-\sigma_{\mu\nu}{\tilde\Gamma}
 \{\sigma^{\mu\nu},v\hspace{-1.7mm}/\}\Gamma)\nonumber\\ &&\times\Bigg(
 \Gamma^\prime\frac{1+v\hspace{-1.7mm}/}{2}{\tilde\Gamma}^\prime\Bigg)~,
\end{eqnarray}
where the subscripts of $\Pi$ denote the figures, $N_c$ is the number of
colors, and ${\tilde\Gamma}\equiv\gamma^0\Gamma^\dagger\gamma^0$.  The
quark condensate ($\langle{\bar q}q\rangle$), gluon condensate
($\langle\alpha_s GG\rangle$), and quark-gluon mixed condensate
($\langle \bar{q}g_s\sigma\cdot G q\rangle$), which are of dimension less
than 6 and the dimension 6 condensate ($\langle{\bar q}q\rangle^2$) are
included.  Note that the term of $\Bigg(\Gamma^\prime
\frac{1+v\hspace{-1.7mm}/}{2}{\tilde\Gamma}^\prime\Bigg)$ appears
commonly.
This is proportional to $\frac{1+v\hspace{-1.7mm}/}{2}$ or 
$\Lambda^{\mu\nu}$ for Eq. (\ref{noderiv}) and (\ref{deriv}), respectively,
which is dropped in Eq. (\ref{mass}).
For Fig. \ref{derivfig}, we obtain the following results, 
\begin{eqnarray}
\frac{1}{\pi}{\rm Im}\Pi_{2a}(\nu)&=&\frac{3N_c!}{4\pi\cdot7!}\nu^7
 {\rm Tr}(\tau^\dagger\tau)(24a^2+40b^2)
 \Bigg(\frac{1+v\hspace{-1.7mm}/}{2}\Bigg)~,\nonumber\\
{\hat B_T^\omega}(\Pi_{2b}(\omega)+\Pi_{2f}(\omega))&=&\frac{N_c!}{2\pi^2}
 {\rm Tr}(\tau^\dagger\tau)\big[\langle{\bar q}q\rangle T^5(16ab)-\langle
 \bar{q}g\sigma\cdot G q\rangle T^3ab\big]\Bigg(\frac{1+v\hspace{-1.7mm}/}
 {2}\Bigg)~,\nonumber\\
{\hat B_T^\omega}\Pi_{2c}(\omega)&=&{\hat B_T^\omega}\Pi_{2g}(\omega)=0~,
 \nonumber\\
{\hat B_T^\omega}\Pi_{2d}(\omega)&=&\frac{\langle\alpha GG\rangle}
 {32\pi^3}T^4{\rm Tr}(\tau^\dagger\tau)(-a^2+b^2)
 \Bigg(\frac{1+v\hspace{-1.7mm}/}{2}\Bigg)~,\nonumber\\
{\hat B_T^\omega}\Pi_{2e}(\omega)&=&-\frac{\langle \bar{q}g\sigma\cdot G
 q\rangle}{4\pi^2}T^3{\rm Tr}(\tau^\dagger\tau)(3ab)
 \Bigg(\frac{1+v\hspace{-1.7mm}/}{2}\Bigg)~,\nonumber\\
\end{eqnarray}
for spin 1/2.
For spin 3/2 case, the results are the same except the factor
$\frac{1+v\hspace{-1.7mm}/}{2}$ replaced by $\Lambda^{\mu\nu}$.

\section{Numerical results}

\begin{figure}
\begin{center}
\epsfig{file=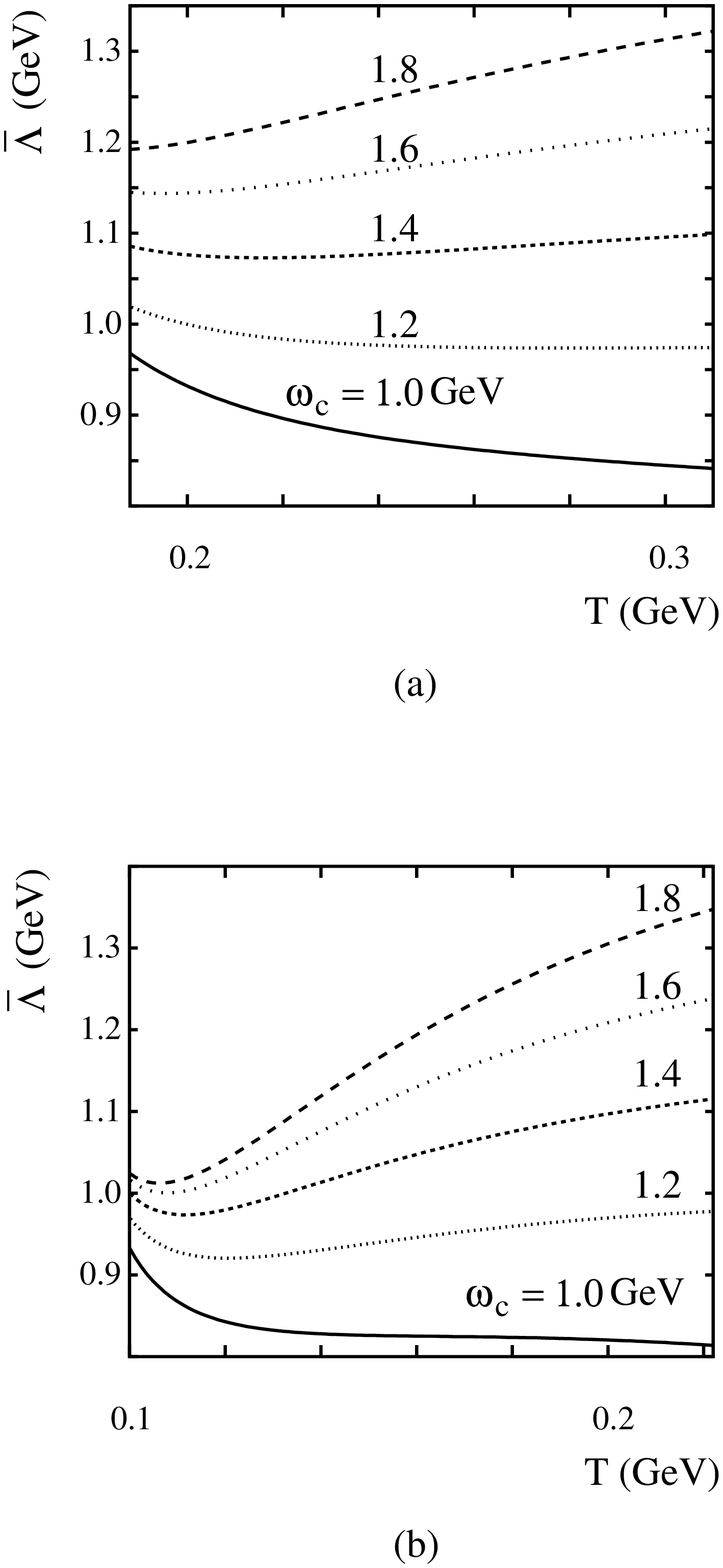, height=20cm}
\end{center}
\caption{}
\label{massfig}
\end{figure}

Fig. \ref{massfig} shows the numerical result of the mass parameter
${\bar\Lambda}$ as a function of the Borel parameter $T$.  The following
values for the condensates have been used, 
\begin{eqnarray}
\langle{\bar q}q\rangle&=&-(0.23~{\rm GeV})^3~,\nonumber\\
\langle\alpha GG\rangle&=&0.04~{\rm GeV}^4~,\nonumber\\
\langle \bar{q}g\sigma\cdot G q\rangle&\equiv&m_0^2\langle{\bar q}q
\rangle~,~~~~~m_0^2=0.8~{\rm GeV}^2~.
\end{eqnarray}
Five curves with various continuum threshold $\omega_c$ are plotted, and
the best value of $\omega_c$ is 
\begin{equation}
\omega_c^{\rm best}=1.40\pm0.20~{\rm GeV}~.
\end{equation}
This number is reasonable considering the quark model results
\cite{Copley}.  The sum rule window is fixed in the following.  The lower
bound of the sum rule window is chosen such that the condensate
contribution is less than 30\%, while the upper bound is determined from
the requirement that the pole contribution is at least 30\%.
The window is determined as 
\begin{eqnarray}
0.188~{\rm GeV}\leq T \leq 0.310~{\rm GeV}&&{\rm for~~baryon~~current
 ~~without~~derivative,} \nonumber\\
0.114~{\rm GeV}\leq T \leq 0.222~{\rm GeV}&&{\rm for~~baryon~~current
 ~~with~~derivative.} 
\end{eqnarray}
From the windows, ${\bar\Lambda}$ is obtained as 
\begin{equation}
{\bar \Lambda}=\left\{ \begin{array}{l}
               1.08^{+0.095}_{-0.104}~{\rm GeV} 
               ~~~~~(\mbox{non-derivative current}),\nonumber\\
               1.06^{+0.090}_{-0.107}~{\rm GeV} 
               ~~~~~(\mbox{derivative current}).
                       \end{array}
               \right.
\end{equation}
For the case of derivative current, the best stability is at $a=1$ and
$b=0$.
From the experimental value of the spin-averaged mass
$\frac{1}{3}(M_{\Lambda_{c1}(\frac{1}{2})}+2M_{\Lambda_{c1}
(\frac{3}{2})})\simeq 2616$ MeV \cite{PDG}, we have
\begin{eqnarray}
m_c&=&\left\{ \begin{array}{l}
         1.54^{+0.104}_{-0.095} ~{\rm GeV} 
         ~~~~~(\mbox{nonderivative current}),\nonumber\\
         1.56^{+0.107}_{-0.090} ~{\rm GeV} 
               ~~~~~(\mbox{derivative current}).
            \end{array} 
       \right.     
\end{eqnarray}
These values are consistent with those obtained from the sum rules for
the heavy mesons \cite{Neubert,Dai0} and the ground state heavy baryons
\cite{Dai}.

\section{Summary and discussion}

We have calculated the mass parameter of the excited $\Lambda_c$ baryons
using QCD sum rules in the framework of HQET.  Two kinds of interpolating
current for the excited heavy baryons are introduced.  It has been
obtained that $\bar{\Lambda}=1.08^{+0.095}_{-0.104}$ GeV for the
non-derivative current and $\bar{\Lambda}=1.06^{+0.090}_{-0.107}$ GeV for
the current with derivative.  These result in $m_c\simeq 1.54\pm 0.1$ GeV
and $1.56\pm 0.1$ GeV, respectively, which are consistent with those from
the analyses of other heavy hadrons.  Continuum threshold is at 
$\omega_c^{\rm best}=1.40\pm0.20~{\rm GeV}$.  Due to HQS, our result is 
also valid for the excited $\Lambda_b^*$ baryons.
\par
This work can be improved through considering $1/m_Q$ and $\alpha_s$
corrections.  The $1/m_Q$ correction which keeps the heavy quark spin
symmetry will affect the spin-averaged mass.  It is at the $15\%$ level
for the ground state $\Sigma_c$ and $\Sigma_c^*$ baryon \cite{Dai}.  The
spin symmetry violating correction contributes to the mass difference of
$\Lambda_{Q1}({\frac{1}{2}}^-)$ and $\Lambda_{Q1}({\frac{3}{2}}^-)$.
While the $\alpha_s$ correction can be significant for the decay constant,
it is expected to be small for the mass parameter ${\bar\Lambda}$ because
of the cancelation in Eq. (\ref{mass}).  

\begin{ack}
We would like to thank Yuan-ben Dai and Ming-qiu Huang for helpful
discussions.  We also thank Chao-shang Huang for informing us that a
similar work in under their analysis \cite{Huang}.  This work was
supported in part by KOSEF through the SRC program of SNU CTP and by 
Korean Research Foundation through the Sughak program and 
1998-015-D00054.
\end{ack}


\newpage
\begin{center}{\large\bf FIGURE CAPTIONS}\end{center}

\noindent
Fig.~1
\\
Diagrams for calculating $\Lambda_{Q1}$ mass with interpolating current
without derivative.  Double line denotes the heavy quark.
\vskip .3cm
\par
\noindent
Fig.~2
\\
Diagrams for calculating $\Lambda_{Q1}$ mass with interpolating current
with derivative.  Double line denotes the heavy quark.
\vskip .3cm
\noindent
Fig.~3
\\
HQET sum rule for the mass parameter ${\bar\Lambda}$ by using baryon
interpolating current with derivative (a) and without derivative (b).
$T$ is the Borel parameter.
\vskip .3cm
\noindent

\newpage

\newpage

\newpage

\end{document}